\documentclass[conference]{IEEEtran}

\usepackage{adjustbox}
\usepackage{float}
\usepackage{graphicx}
\usepackage{framed,multirow}
\usepackage{subcaption}
\usepackage{subfig}
\usepackage{xcolor}
\usepackage{booktabs}
\usepackage{makecell}
\usepackage{amsmath}
\usepackage{soul}
\usepackage[normalem]{ulem}

\usepackage[noadjust]{cite}

\begin{document}


\title{Leveraging Multiphase CT for Quality Enhancement of Portal Venous CT:  \\ Utility for Pancreas Segmentation}

\author{%
  \IEEEauthorblockN{%
    \parbox{\linewidth}{\centering
      Xinya Wang,
      Tejas Sudharshan Mathai,
      Boah Kim,
      Ronald M. Summers
    }%
  }%
  \IEEEauthorblockA{%
     Radiology and Imaging Sciences, National Institutes of Health Clinical Center, Bethesda, MD, USA \\
    Email: tejas dot mathai at nih dot gov%
  }%
}
\maketitle

\begin{abstract}

\noindent
Multiphase CT studies are routinely obtained in clinical practice for diagnosis and management of various diseases, such as cancer. However, the CT studies can be acquired with low radiation doses, different scanners, and are frequently affected by motion and metal artifacts. Prior approaches have targeted the quality improvement of one specific CT phase (e.g., non-contrast CT). In this work, we hypothesized that leveraging multiple CT phases for the quality enhancement of one phase may prove advantageous for downstream tasks, such as segmentation. A 3D progressive fusion and non-local (PFNL) network was developed. It was trained with three degraded (low-quality) phases (non-contrast, arterial, and portal venous) to enhance the quality of the portal venous phase. Then, the effect of scan quality enhancement was evaluated using a proxy task of pancreas segmentation, which is useful for tracking pancreatic cancer. The proposed approach improved the pancreas segmentation by $\sim$3\% over the corresponding low-quality CT scan. To the best of our knowledge, we are the first to harness multiphase CT for scan quality enhancement and improved pancreas segmentation. 


\end{abstract}



\section{\textbf{Introduction}}

\noindent
Multi-phase computed tomography (CT) studies are routinely obtained in clinical practice to visualize and diagnose a variety of conditions, such as cirrhosis \cite{Lee2022}, coronary artery disease \cite{sandfort_denoising_2024}, diabetes \cite{Tallam2022}, and pancreatic cancer \cite{Prasad2025_pancreas,suri_comparison_2024}. Currently, radiologists utilize multiple complementary CT phases, such as non-contrast (native), arterial, portal venous (PV), and delayed sequences, to render a diagnosis. Among the individual CT phases, the portal venous phase is valuable for lesion detection \cite{Lee2022} and vessel segmentation \cite{jaouen_self_2023}. However, the quality of acquired CT scans can vary across different institutions in the United States. This is because there are many CT scanners from different manufacturers and diverse CT exam protocols in use. Patients are also scanned with low radiation doses, and the scans can often be affected by metal artifacts or motion. Abdominal CT volumes are also reconstructed with non-isotropic voxel resolutions ranging from 0.5-5 mm in the in-plane and through-plane directions. This leads to diverse appearances of organs and structures in CT. 

To mitigate these effects, numerous approaches \cite{Zhao2024_review,Lei2024_review,Li2023_review,Zhang2021_SADIR_sinogramDomain,zohair2015latest,Sandfort2019,sandfort_denoising_2024,jaouen_self_2023} in the past have focused on CT-based denoising and CT super-resolution with promising clinical applications. Several works have mostly focused on improving the quality of a single CT phase~\cite{zhao_smore:_2021, ChenISBI2017,LUO2022102335,Wang_2023ctformer,YangMICCA2024,yang2024amir}. For example, Zhang \emph{et al.}~\cite{Zhang2021_SADIR_sinogramDomain} combined a super-resolution model in the sinogram domain and an image deblurring model in the image domain for portal venous CT super-resolution. Jaouen \emph{et al.}~\cite{jaouen_self_2023} studied a self super-resolution network to improve segmentation accuracy of hepatic vessels from abdominal CT scans. Sandfort \emph{et al.}~\cite{sandfort_denoising_2024} denoised dose modulated coronary CT angiography scans from multiple time points (longitudinal scans) using a 3D U-Net model. However, to our knowledge, there is no prior approach that leverages the multiple phases inherently available within a CT study for scan quality enhancement. 

\begin{figure*}[!t]   
    \centering
    \includegraphics[width=0.914\textwidth]{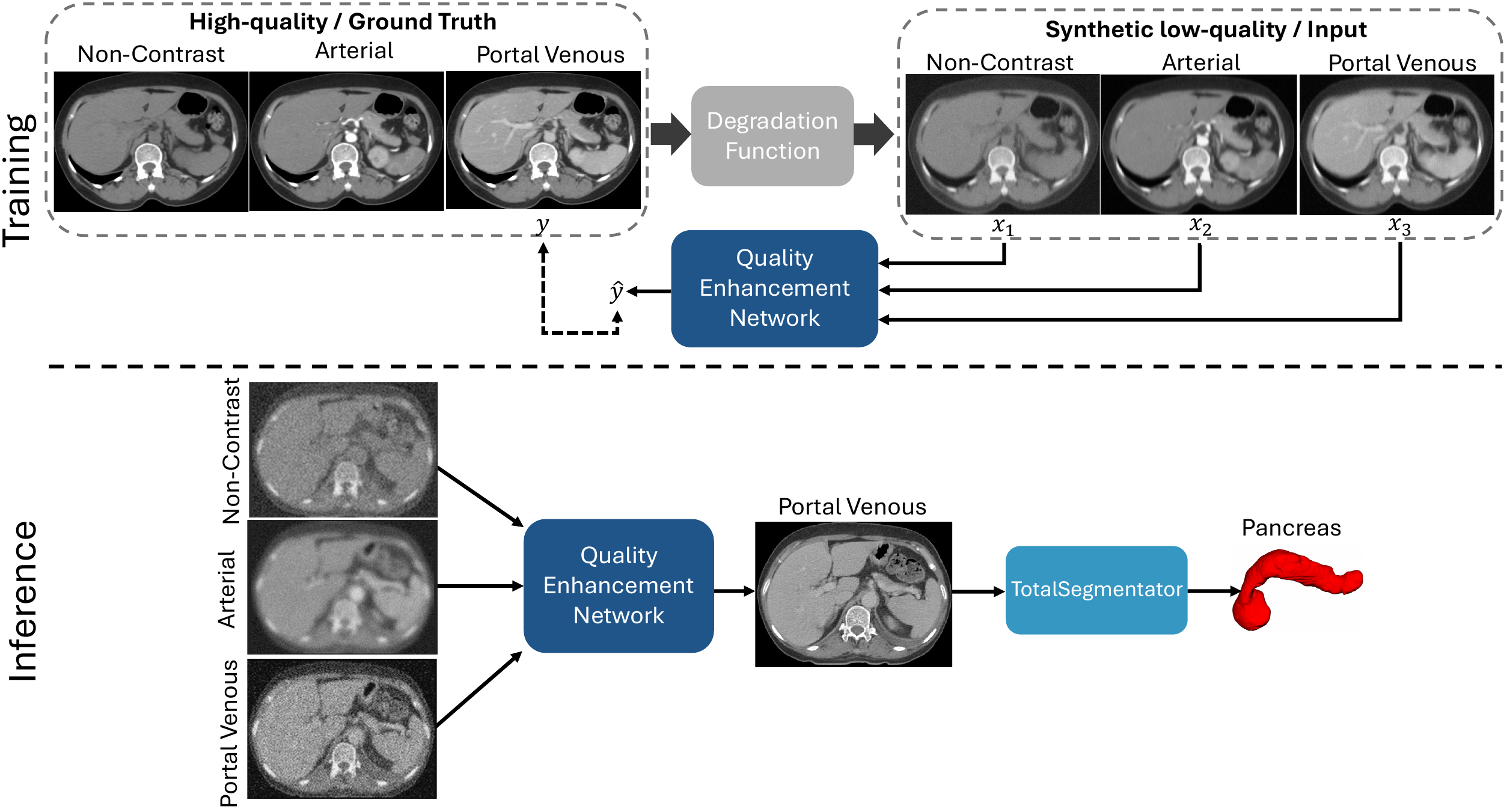}
    \caption{The overall framework for portal venous scan quality improvement is shown. Three degraded and low-quality CT phases (non-contrast, arterial, and portal venous) were used to train a 3D-PFNL quality enhancement network. At test time, the model generated a high-quality portal venous CT volume, given the three inputs. The effect of scan quality enhancement was assessed with the public TotalSegmentator tool that segmented the pancreas in a proxy segmentation task.}
    \label{fig_pipeline}
\end{figure*}

In this proof-of-concept study, we utilized multiphase CT scans to enhance the quality of the portal venous phase CT at the level of abdomen. As shown in Fig.~\ref{fig_pipeline}, the proposed approach takes three degraded (low-quality) CT phases (non-contrast, arterial, and portal venous) as inputs and produces a high-quality portal venous CT. We extended the progressive fusion and non-local (PFNL) model \cite{YiPFNL2019} into 3D, and trained it with the loss function that combined an ${L}_{1}$ reconstruction loss with a 3D Sobel edge-based loss. To determine the effect of quality enhancement, a proxy pancreas segmentation task using the public TotalSegmentator tool~\cite{wasserthal_totalsegmentator:_2023} was performed. Our results demonstrated that leveraging multiphase CT for quality enhancement of the portal venous phase improved the pancreas segmentation by $\sim$3\%. To the best of our knowledge, we are the first to harness multiphase CT for scan quality enhancement and improved pancreas segmentation performance.

\section{\textbf{Methods}}

\subsection{\textbf{Patient Sample}}

\noindent
In this retrospective study, the publicly available VinDr-Multiphase dataset~\cite{dao_phase_2022} was used. It contains $265$ abdominal CT studies from $265$ patients with $1188$ CT series acquired between 2015 and 2020. The dataset was originally published for CT phase classification. The studies were retrospectively selected from the PACS systems of two Vietnamese hospitals. Volume dimensions ranged from 512 $\times$ 512 $\times$ (30\text{-}2350) voxels and the spacing ranged between 0.5-5 mm. Each scan was provided one of four labels: non-contrast, arterial, portal venous and others (e.g., delayed phase). In this work, only studies containing the following three CT phases were chosen: (1) non-contrast, (2) arterial and (3) portal venous (PV). Studies that did not contain these three phases were excluded. The final dataset had a total of 168 studies from 168 patients, and was split at the patient level to avoid data leakage into training (80\%, n = 148), validation (10\%, n = 16) and testing (10\%, n = 16) data subsets. 

\subsection{\textbf{Reference Standard}}

\noindent
Segmentation of the pancreas is clinically meaningful for assessing pancreatic pathologies~\cite{Tallam2022,suri_comparison_2024}. In this work, the pancreas was manually annotated in the original portal venous phase CT scans in the test data subset (n = 16 scans). This step was required to assess the quality enhancement of the PV CT scan using a proxy segmentation task. A two-stage approach for annotation of the pancreas was followed. First, a public segmentation model~\cite{zhang2024large} was used to automatically segment the pancreas. Then, these automatic annotations were manually corrected and verified by a board-certified radiologist with 30+ years of experience. 


\subsection{\textbf{Data Preprocessing}}

\noindent
First, the non-contrast and arterial phases were resampled to have a consistent voxel resolution as the PV phase CT. Next, the PV phase was set as the reference, and the non-contrast and arterial phases were co-registered to it using ANTS~\cite{avants_reproducible_2011}. Following this step, the three co-registered CT volumes were cropped to the abdomen, and windowed with the level of 50 HU and a width of 450 HU. To generate low-quality CT scans, a multi-stage degradation process was adopted to simulate real degradations to the CT volumes~\cite{wang_real-esrgan:_2021}. This was because the VinDr dataset did not contain paired high- and low- quality CT scans. Thus, it was necessary to degrade the original CT scans, such that the model could enhance and restore the quality of the CT volumes. In each degradation stage, a random combination of blurring (isotropic Gaussian and anisotropic Gaussian), noise (Gaussian and Poisson) and resizing operations (at the scale factor $4$) was chosen to emulate intensity degradations. Based on prior empirical work~\cite{wang_real-esrgan:_2021}, a second-order degradation process was adopted for a good balance between simplicity and effectiveness. After degradation, synthetically degraded CT volumes of low quality were used for training.

\subsection{\textbf{Model for Portal Venous Phase Quality Enhancement}}

\noindent
Fig.~\ref{fig_pipeline} shows an overview of the approach. In the training stage, the original CT scans were the reference high-quality images. The quality enhancement network took the three low-quality CT scans and was trained to recover the high-quality portal venous phase $\hat{y}=R(x_{0},x_{1},x_{2})$, where $x_{0},x_{1},x_{2}$ denote the non-contrast, arterial and portal venous phases, respectively. The quality enhancement network $R$ was based on the progressive fusion and non-local (PFNL) architecture \cite{YiPFNL2019}. It was initially designed for 2D video super-resolution, but it was extended into 3D in this work. Our 3D-PFNL model leveraged multiphase CT scans for the quality enhancement of the portal venous phase. The objective function used for training the model comprised of an $L_1$ reconstruction loss term and an $L_1$ Sobel-edge reconstruction loss term:

\begin{equation}
    \mathcal{L}(y,\hat{y})=\mathcal{L}_{1}(y,\hat{y})+\lambda \mathcal{L}_{1}(S(y),S(\hat{y})),
\end{equation}
in which $S$ is the 3D Sobel operator to extract the edges of structures. The intent of this edge reconstruction loss was to allow the network to focus on the boundaries of the anatomical structures in CT scans. This is particularly useful for both contrast and non-contrast CT scans where partial volume averaging can affect the delineation of organs, such as the pancreas. The parameter $\lambda$ was empirically set to 0.7 based on prior work. 

For the training phase, data augmentation was carried out through random rotations of 90, 180, or 270 degrees. The model was trained for a total of 1000 epochs with a batch size of 8, a learning rate of $10^{-3}$, and an Adam optimizer. All experiments were conducted using the Pytorch library in Python with a single NVIDIA A100 GPU. During inference, the degraded CT scans were restored to their high-quality counterpart. 

\subsection{\textbf{Pancreas Segmentation}}

\noindent
In this work, the publicly available TotalSegmentator (TS) \cite{wasserthal_totalsegmentator:_2023} tool was used to segment the pancreas in the restored PV CT. TS can segment 117 structures in CT to date, and it was mostly trained on contrast-enhanced CT scans. While it can be used to segment all 117 structures in the VinDr dataset, their segmentations would need to be manually verified. To reduce the annotation verification burden, the focus of the current work was solely on the pancreas due to its clinical utility. If the quality enhancement process fails, then the segmentation of the pancreas by TS would be poor. Thus, this proxy segmentation task determined the quality enhancement effect on the portal venous phase.

\subsection{\textbf{Experiments}}
\noindent
The 3D-PFNL model was compared against an extended residual channel attention (3D-RCAN) network~\cite{zhang2018rcan}, which is a single image super-resolution technique. 3D-RCAN was only trained with the portal venous phase CT from the VinDr dataset, whereas 3D-PFNL was trained on all three phases. For the proxy pancreas segmentation task, TS was run on the original PV CT scans, the low-quality (LQ) CT scans, and also on the PV CT scans restored in quality by 3D-RCAN and 3D-PFNL. 

\subsection{\textbf{Statistical Analysis}}
\noindent
Image quality enhancement metrics, such as PSNR and SSIM, were calculated to measure the effect of using multiphase CT for restoration of the PV phase. Similarly, Dice similarity coefficient (DSC) and normalized surface distance (NSD) were computed to quantify the pancreas segmentation performance by TS on the restored PV phase scans. As the Dice and NSD scores were not normally distributed, a paired Wilcoxon signed-rank test was used for statistical testing. A p-value $<.05$ was considered statistically significant.

\section{\textbf{Results}}

\subsection{\textbf{Quality Enhancement Results}}

\begin{figure}[!t]   
    \centering
    \includegraphics[width=0.46\textwidth]{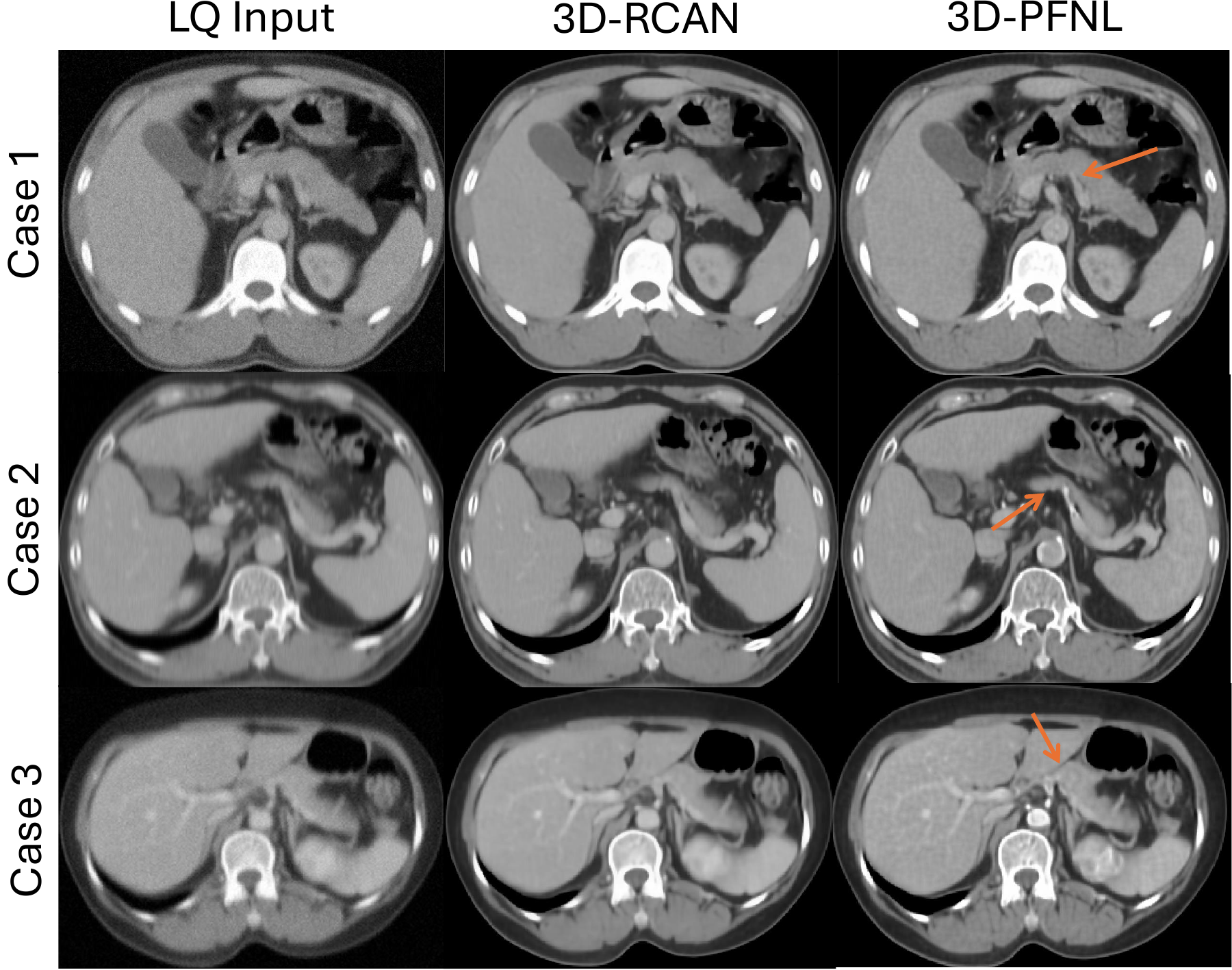}
    \caption{Qualitative examples of low-quality (LQ) portal venous (PV) CT, 3D-RCAN restored PV CT and 3D-PFNL restored PV CT, respectively. Orange arrows indicate the sharper and clearer edges of the pancreas by the proposed 3D-PFNL quality enhancement approach.}
    \label{fig_recon}
\end{figure}

\noindent
As shown in Table~\ref{t_recon}, both 3D-PFNL and 3D-RCAN models enhanced the quality of the LQ PV CT. There was very little difference ($<$ 1\%) between the 3D-PFNL and 3D-RCAN models. As seen in Fig.~\ref{fig_recon}, the proposed 3D-PFNL method recovered sharper edges of the pancreas, whereas the 3D-RCAN model over-smoothed the edges of the pancreas. Of note, the edge reconstruction loss used in the 3D-PFNL model shifted the focus from perceptual quality to structural enhancement. This is in contrast to existing restoration models~\cite{zhang2018rcan, YangMICCA2024, yang2024amir} that are optimized by the $L_1$ reconstruction loss alone. However, an increase in quality does not necessarily mean that it can facilitate downstream tasks, such as segmentation. 

\begin{table}[!t]
\centering
\caption{Comparison of quality enhancement networks evaluated on the test data subset.}
\begin{tabular}{ccc}
\hline
Method & PSNR & SSIM \\ \hline
LQ Input    & 23.82     &  0.4591    \\
3D-RCAN     & 28.77     &  0.8794    \\
3D-PFNL     & 28.16     &  0.8733    \\ \hline
\end{tabular}
\label{t_recon}
\end{table}

\subsection{\textbf{Segmentation Results}}
\begin{figure}[!t]
	\begin{minipage}[t]{0.49\linewidth}
		\centering
		\includegraphics[width=1.7in]{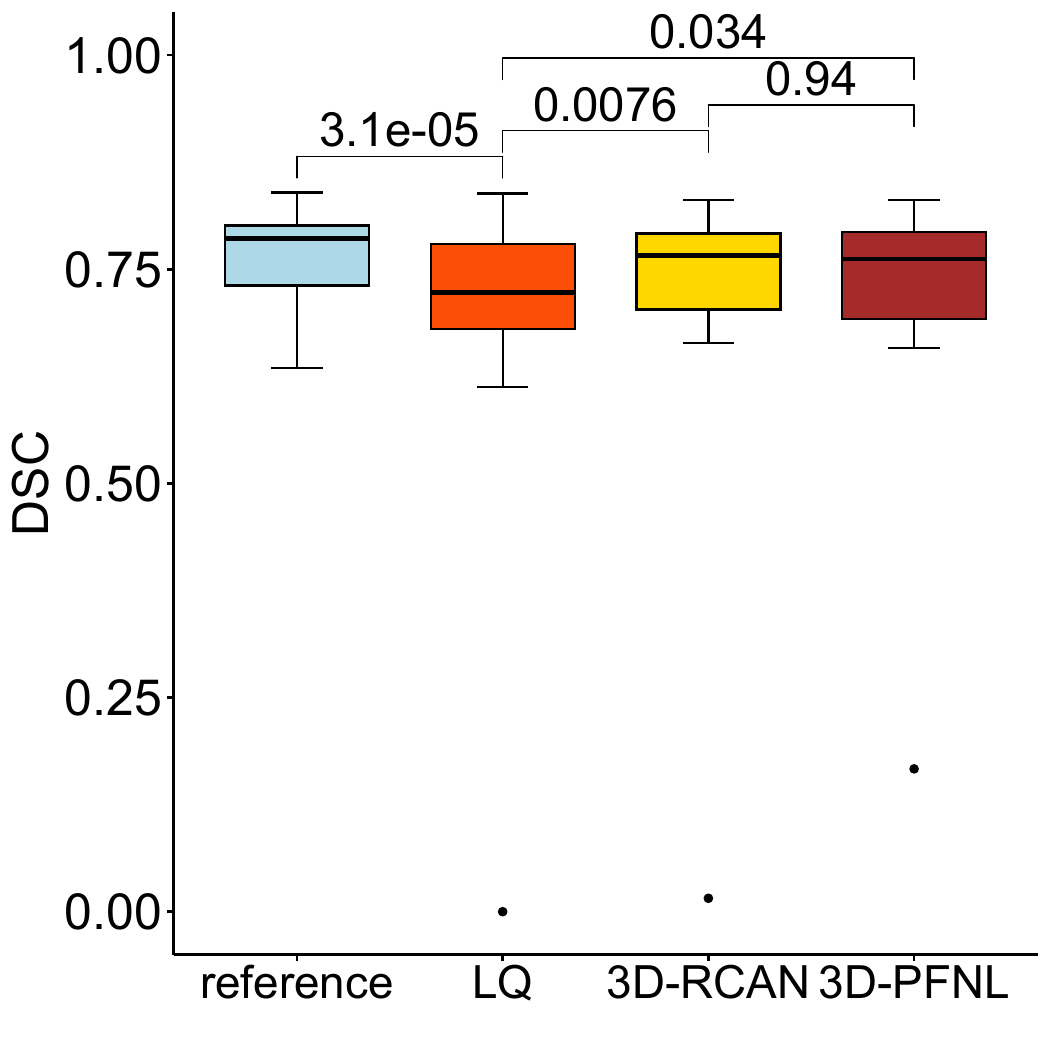}
            \subcaption{Dice score}
	\end{minipage}
	\begin{minipage}[t]{0.49\linewidth}
		\centering
		\includegraphics[width=1.7in]{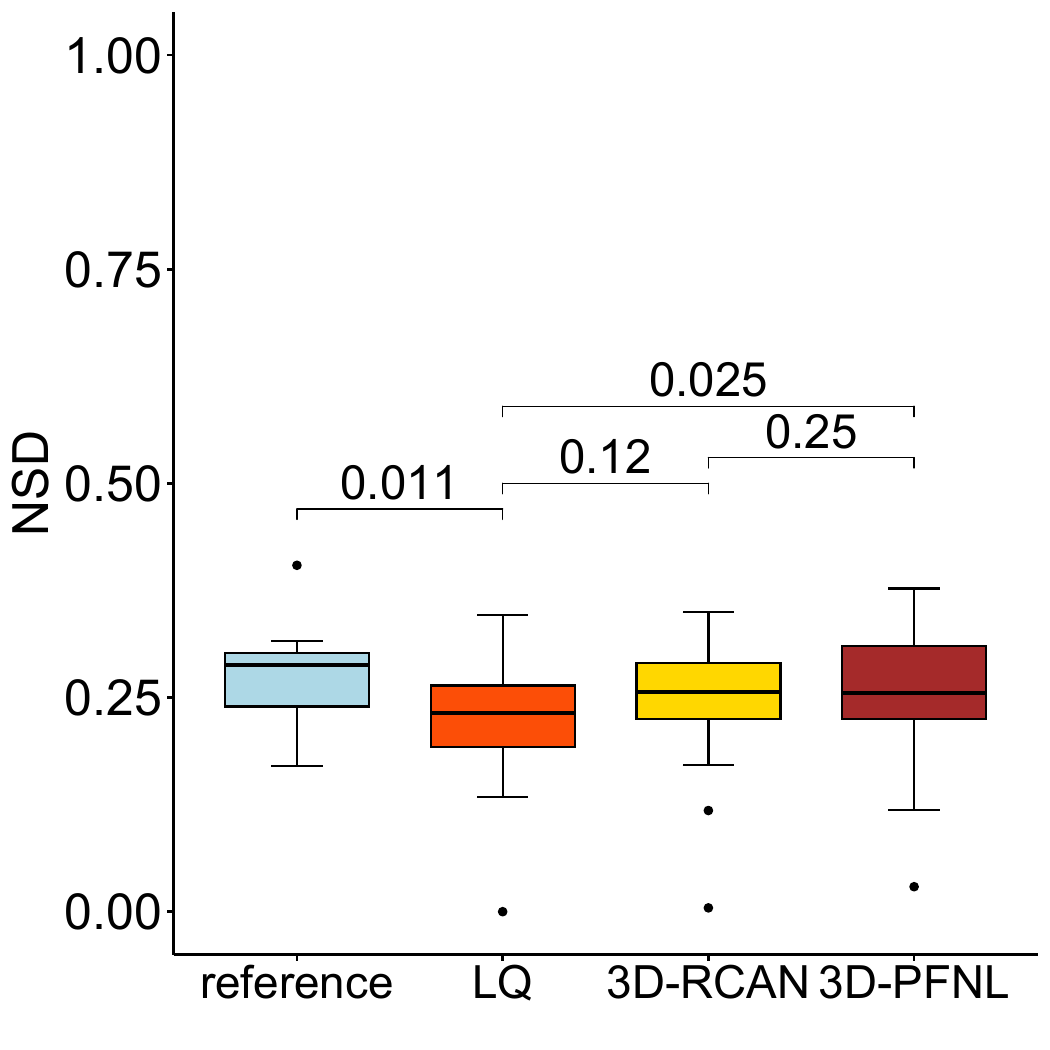}
            \subcaption{Normalized Surface Distance}
	\end{minipage}
	\caption{Box plots of (a) Dice score (DSC) and (b) Normalized surface distance (NSD) for the pancreas segmentation by TotalSegmentator on reference (original) portal venous (PV) CT scan, low-quality (LQ) PV CT, 3D-RCAN restored PV CT and 3D-PFNL restored PV CT, respectively.}
	\label{fig_box}
\end{figure}

\noindent
Fig.~\ref{fig_box} shows the distribution of Dice scores and NSD for pancreas segmentation by TS on the reference (original) PV CT, low-quality (LQ) PV CT, 3D-RCAN restored PV CT, and 3D-RFNL restored PV CT, respectively. From Table~\ref{t_seg}, the reference corresponded to the baseline performance of TS on the original PV CT for pancreas segmentation. However, when TS was executed on the degraded low-quality PV CT, the Dice score and NSD dropped to 68.9 $\pm$ 19.3 and 22 $\pm$ 8, respectively. After PV phase quality enhancement with 3D-RCAN, the Dice score and NSD improved by approximately 1.9\% and 2\%, respectively. Moreover, quality enhancement with 3D-PFNL further improved the Dice score by approximately 2.8\% and NSD by 3\%. Moreover, 3D-PFNL attained a lower standard deviation in Dice scores compared to both LQ and 3D-RCAN. The confidence interval of Dice score and NSD obtained by 3D-PFNL are [0.6299, 0.7725] and [0.2066, 0.2869] respectively.

\begin{figure}[!t]   
    \centering
    \includegraphics[width=0.45\textwidth]{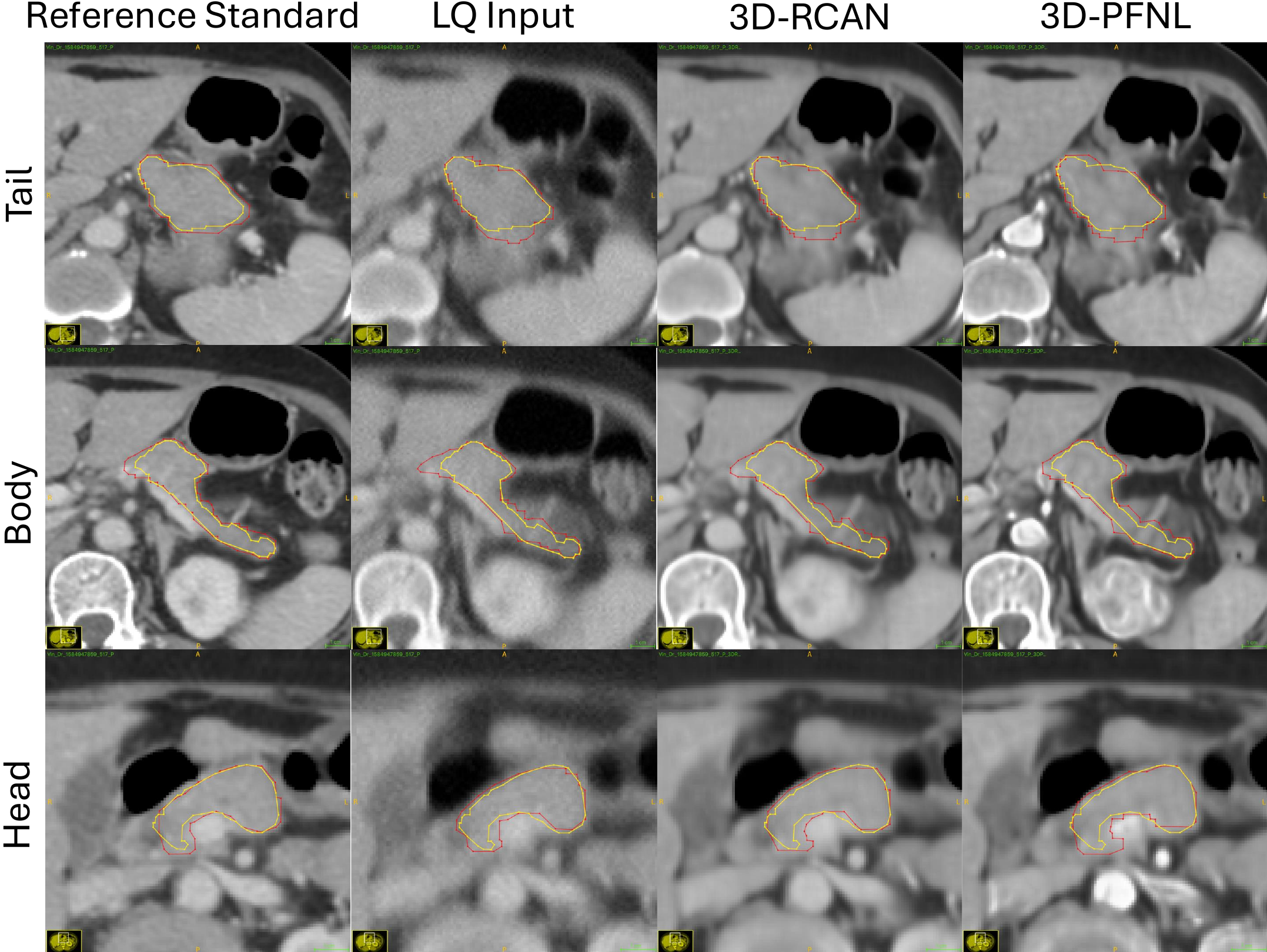}
    \caption{Segmentation results of TotalSegmentator (TS) on the reference (original) portal venous (PV) CT scan, low-quality (LQ) PV CT, 3D-RCAN restored PV CT and 3D-PFNL restored PV CT, respectively. Manual pancreas annotations are in yellow, while the automated segmentations by TS are in red.}
    \label{fig_seg}
\end{figure}

Compared against LQ, results from 3D-PFNL on the restored PV CT were statistically significant for both Dice score (p = .034) and NSD (p = .025). The Dice score on the restored PV CT was also significant for 3D-RCAN (p = .007). Notably, TS failed to segment the pancreas in one LQ PV CT. After quality enhancement with 3D-PFNL, TS segmented the pancreas in the same volume. Fig.~\ref{fig_seg} shows an example. The pancreatic body (second row in Fig.~\ref{fig_seg}) and tail (first row) tended to be over-segmented, while the pancreas head (third row) was under-segmented. After quality enhancement, TS segmented most of the pancreas with minor errors. 



\begin{table}[t]
\centering
\caption{Results of pancreas segmentation by TotalSegmentator on reference (original) portal venous (PV) CT scan, low-quality (LQ) PV CT, 3D-RCAN restored PV CT and 3D-PFNL restored PV CT, respectively. DSC: Dice Score. NSD: Normalized Surface Distance.}
\begin{tabular}{ccc}
\hline
\multirow{2}{*}{Method} & \multicolumn{2}{c}{Metrics} \\
                                  & DSC (\%)      & NSD (\%)      \\ \hline
Reference                         & 76.5 $\pm$ 5.70        & 27.1 $\pm$ 6.10       \\
LQ Input                         & 68.9 $\pm$ 19.3        & 22.0 $\pm$ 8.00       \\
3D-RCAN                      & 70.8 $\pm$ 19.1        & 24.0 $\pm$ 8.60        \\
3D-PFNL                      & 71.7 $\pm$ 15.6        & 25.0 $\pm$ 8.50        \\ \hline
\end{tabular}
\label{t_seg}
\end{table}

\section{\textbf{Discussion}}

\noindent
In this work, a 3D-PFNL model was proposed to leverage three multiphase CT scans (non-contrast, arterial, and portal venous) to enhance the quality of the PV phase CT. To our knowledge, we are the first to propose the use of multiphase CT for image quality enhancement of a specific CT phase, such as portal venous CT. Compared to the 3D-RCAN model, which used only the PV phase CT for training, the multiphasic 3D-PFNL model recovered clear and sharper boundaries of target organs. Note that the proposed approach was designed to focus more on the boundaries, rather than being tailored to specifically improve the visual quality of the pancreas. This enables our work to be generalizable to all organs and structures in the body. Additionally, the proposed approach can also be used to enhance the quality of other phases, such as non-contrast CT, which has been shown to be useful for opportunistic screening~\cite{Tallam2022}. 

Both 3D-PFNL and 3D-RCAN performed comparably, but this did not imply facilitation of downstream tasks~\cite{Michaeli2018,yang2019deep}, such as segmentation. Thus, the proxy task of pancreas segmentation was used to verify the effect of image quality improvement. TotalSegmentator segmented the pancreas in the restored PV CT scans from both models (p $<$ .05) in contrast to the low-quality PV CT. In contrast to 3D-RCAN, 3D-PFNL enabled TS to close the gap in segmentation performance (via lower Dice standard deviations) between the original and restored PV CT. As TotalSegmentator was trained on a different dataset, the Dice scores on the out-of-distribution VinDr dataset were below 80\%, which is in line with prior literature on pancreas segmentation \cite{Prasad2025_pancreas,suri_comparison_2024}. 

However, there was no significant difference in the performance between the two models. This can be attributed to the small number of patients (n = 16) in the test data subset, and it is the major limitation of our work. Another limitation is the introduction of artifacts on the portal venous phase CT. 3D-PFNL attempted to fuse information from the three phases to reconstruct a high-quality PV phase. As shown in case 3 of Fig.~\ref{fig_recon} and Fig.~\ref{fig_seg}, the contrast of the aorta was enhanced in the portal venous phase, but it is not a feature of this phase. Furthermore, low-quality PV CT was simulated through intensity-based degradations, and the proposed model was trained and evaluated on the same VinDr dataset. The generalizability of the model to external datasets (e.g., low-dose CT) is currently undetermined. In the future, validation of the approach on a larger sample size and with additional organs and structures is necessary. 




\clearpage

\section{ACKNOWLEDGMENTS}      

\noindent
This work was supported by the Intramural Research Program of the NIH Clinical Center (project number 1Z01 CL040004). The research used the high-performance computing facilities of the NIH Biowulf cluster.

\section{Compliance with Ethical Standards}

This study used a publicly available anonymized dataset. Institutional Review Board approval was not required.

\bibliographystyle{IEEEtran}
\bibliography{references}

\end{document}